\documentclass[10pt]{article}
\usepackage{graphicx}
\begin{document}

\title{
 {\bf \Large Effect of Variable Surrounding on Species Creation}
}

\author{Aleksandra Nowicka$^1$,
Artur Duda$^2$, and  Miros{\l}aw R. Dudek$^3$
}

\date{}

\maketitle

\begin{center}
{$^1$ \it Institute of Microbiology, University of Wroc{\l}aw,
ul. Przybyszewskiego 63/77\\
54-148 Wroc{\l}aw, Poland}\\
~\\
{$^2$ \it Institute of Theoretical Physics, University of Wroc{\l}aw,
        pl. Maxa Borna 9\\
 50-204 Wroc{\l}aw, Poland \\
 ~\\
~\\
$^3$  \it Institute of Physics, Zielona G{\'o}ra University,
 65-069 Zielona G{\'o}ra, Poland
}
 \end{center}

\normalsize

\begin{center}
{\bf Abstract}
\end{center}
{\small
We construct a model of speciation from evolution
in an ecosystem consisting of a limited amount of
energy recources.
The species posses genetic information, which is
inherited according to the rules of the Penna model
of genetic evolution. The increase in number of
the individuals of each species depends on the quality
of their genotypes and the available energy resources.
The decrease in number of the individuals
results from the genetic death or
reaching the maximum age by the individual.
The amount of energy resources is
represented by a solution of the
differential logistic equation, where the growth rate
of the amount of the energy resources has been modified
to include the number of individuals from all species
in the ecosystem under consideration.

The fluctuating surrounding is modelled
with the help of the function $V(x,t)=\frac{1}{4} x^4 +
\frac{1}{2} b(t) x^2$, where $x$ is representing
phenotype  and the coefficient $b(t)$
 shows the  $cos(\omega t)$ time dependence.
 The closer the value $x$ of an individual
 to the minimum of $V(x,t)$ the better adapted its genotype to the surrounding.
We observed that the life span of the organisms strongly
depends on the value of the frequency  $\omega$. It becomes the shorter
the often are the changes of the surrounding.
However, there is a tendency
that the species which have a higher value
$a_R$ of the reproduction age win the competition with the other species.

Another observation is
that small evolutionary changes of the inherited genetic information
lead to spontaneous bursts of the evolutionary activity when many
new species may appear in a short period.
}

\section{Introduction}
There have appeared many papers and books on
 the dynamics in ecological systems. Many of them start from
the Lotka-Volterra model \cite{lotka1,lotka2}  which describes
populations in competition.
 Usually the populations are represented by the various types of the
predator-prey systems. They may exhibit
many interesting features such as chaos (e.g. the recent papers on the topic
\cite{chaos} -- \cite{chaos2})
and phase transitions (e.g. \cite{chaos,Rozenfeld,Artur}).
The possible existence of chaos became
evident since the work of May   \cite{may1,may2}. However, usually the studies
concerning chaos in biological populations do not contain the discussion of the
role of the inherited genetic information in it.
The discussion of a predator-prey model
with genetics has been started by Ray et al. \cite{l_Ray1},
who showed that the system  passes from the oscillatory
solution of the Lotka-Volterra equations into a steady-state regime, which
exhibits some features of self-organized criticality (SOC).
Our study \cite{Artur} on the topic
was the Lotka-Volterra dynamics of two
competing populations, prey and predator, with the genetic information
inherited according to the Penna  model \cite{l_penna1} of genetic evolution
and we showed that during time evolution, the populations can experience
a series of dynamical phase transitions which are connected with the
different types of the dominant phenotypes present in the populations.
Evolution is understoood as the interplay of the two processes: mutation,
 in which the DNA of the organisms experiences small chemical changes, and
 selection, through which the better adapted organisms
 have more offsprings than the others. The problem  of speciation
from evolution has been studied recently by McManus et al.
\cite{l_Ray2} in terms of a microscopic model. They confirmed that
the mutation and selection
are sufficient for the appearance of the speciation.
We followed their result and in this study we consider a closed ecosystem
with a variable number of species competing for the same energy resources.
The total energy of the ecosystem cannot exceed the value $\Omega$ and
every individual costs a respective amount of energy units.
We show that the small evolutionary changes of the inherited
genetic information result in the bursts of evolutionary activity
during which new species appear. Some of the species become better adapted
to the fluctuating surrounding and they
win the competition for the energy resources.

Our model belongs to the class of Lotka-Volterra
systems describing one prey (represented by a self-regenerating energy
resources of the ecosystem) and a variable number of the predators
competing for the same prey.

\section{Evolution of energy resources}
All species in the ecosystem under consideration
use the same energy resources.
In the model, the number $N_E(t)$ of the energy units available for the species
satisfies the differential equation

\begin{equation}
       \frac{d N_{E}(t)}{d t} = \varepsilon_E N_{E}(t)
       (1- \gamma_E \frac{N(t)}{\Omega}) (1-\frac{N_{E}(t)}{\Omega})
\label{eq_LV1}
\end{equation}

\noindent
with the initial condition

\begin{equation}
\frac{1}{\Omega} N_E(t_0)={\alpha},
\label{eq_warpocz}
\end{equation}
\noindent
where $N(t)$ represents the total
number of all individuals in the ecosystem (from all species),
$\varepsilon_E$ is the regeneration rate coefficient of the energy resources,
$\gamma_E$ represents relative decrease of $\varepsilon_E$
caused by the living organisms. The equation means that regeneration of the
energy resources in the ecosystem
takes some time and its speed depends
on the number of the living organisms.
The above equation also ensures that the size of the
species cannot be too big and the number
of the species coexisting in the same ecosystem is limited. Otherwise
they would exhaust all the energy resources
of the ecosystem necessary for life processes.

In the case when $N(t)=0$, i.e., if there are no living organisms in the ecosystem,
the Eq.\ref{eq_LV1} reduces to the well known
logistic differential equation with the
following analytical solution \cite{l_logistic}

\begin{equation}
\frac{1}{\Omega}N_E(t)=\frac{\alpha}{\alpha+(1-\alpha) \exp {-\varepsilon_E (t-t_0)}},
\label{rozw1}
\end{equation}

\noindent
and the initial condition  Eq.\ref{eq_warpocz}.

We adapt the above solution into our model (Eq.\ref{eq_LV1}).
To this end we assume that the
individuals from all species in the ecosystem under consideration
may reproduce only at discrete time $t=0, 1, 2, 3, \ldots$ (otherwise $N(t)=const$),
whereas $N_E(t)$
remains a continuous function of $t$ between these discrete time values.
Say, if there is $N(t_0)$ individuals at the discrete time value $t=t_0$ then the
value $N(t)$ remains constant ($N(t)=N(t_0)$)  in the whole
time interval $[t_0,t_0+1)$. The analytical solution
of Eq.\ref{eq_LV1} in this time interval is the following

\begin{equation}
\frac{1}{\Omega}N_E(t)=\frac{\alpha}{\alpha+(1-\alpha)
\exp {-\varepsilon_E (1- \gamma_E \frac{N(t)}{\Omega})(t-t_0)}} ~~for~~ t<t_1,
\label{rozw2}
\end{equation}

\noindent
where the initial condition is represented by Eq.\ref{eq_warpocz} and $t_1=t_0+1$.
At time $t=t_1$ the species reproduce themselves and a  new value, $N(t_1)$,
becomes the initial condition for the next time interval, $[t_1,t_1+1)$.
The numbers $N(t)$ ($t=0, 1, 2, 3, \ldots$) result from the computer simulation
of mutation and selection applied to the species in the ecosystem
according to the Penna  model \cite{l_penna1} of genetic evolution.
The Penna model of evolution represents evolution of bit-strings (genotypes),
where the different bit-strings replicate with some rates and they mutate.
Hence, in our model we have two types of dynamics, the continuous
one for the number $N_E(t)$ of energy units
and the discrete one for genetic evolution of bit-strings.

\section{Species evolution}

We restrict ourselves to diploid organisms and
we follow the biological species concept
 that the individuals belonging to different species cannot reproduce
themselves, i.e., they represent genetically isolated groups.
The populations of each species are
 characterized by genotype,
phenotype and sex.
Once the individuals
are diploid organisms
there are two copies of each gene (alleles) in their genome - one
member of each pair is contributed by each parent.
In our case, the genotype is determined by $2L$ alleles (we have chosen $L=16$ in
computer simulations)
located in two chromosomes.
We agreed that the first $L'$ sites  in the chromosomes
represent the
housekeeping genes \cite{l_cebrat0}, i.e.
the genes which are necessary during the whole life of
every organism.
 We assumed that there  also exist  $L-L'$ additional
"death genes", which are switched on at
a specific age $a=1,2, \ldots, L-L'$ of living
individual. The idea of the chronological genes has been
borrowed from the Penna model
\cite{l_penna1,l_penna2,l_book} of biological ageing.
The term "death gene" has been introduced by Cebrat
\cite{l_cebrat} who discussed the biological meaning of the
 genes which are chronologically
switched on in the Penna model.
The maximum age of individuals   is set to $a=L-L'$. It is the same
for all species.
All species have also the same number $L'$ of the housekeeping genes
and the same number $L-L'$  of the chronological genes.
However, they differ
in the reproduction age $a_R$, i.e., the age at which the individual can
produce the offsprings.
The individuals can
die earlier due to inherited defective genes. We assume, that
it is always the case, when an individual reaches the age $a$
and in its history until the age $a$ there have appeared three
inactive genes in the genotype (inactive gene means two inactive alleles).

We make a simplified assumption, that all organisms fulfill the same life functions
and each function of an organism (does not matter what species)
has been coded with a bit-string
consisting of $16$ bits generated with the help of a computer
random number generator.
We have decided on  $L$
different functions
and the corresponding
bit-strings represent the patterns for the genes.
The genes of all species
are represented by bit-strings which
differ from the respective pattern by a Hamming
distance $H \le 2$. Otherwise the bit-strings do not represent the genes.
In order to distinguish the species, we have
introduced the concept of an ideal predecessor, called Eve, who uniquely
determines all individuals belonging to the particular species.
Namely, the individuals
have genes which may differ from the genes of Eve only by a Hamming
distance $H \le 1$. Genes, which differ from the genes of Eve by a Hamming
distance $H > 1$ and, simultaneously, which differ from
the pattern by a Hamming distance $H \le 2$
represent mutated genes. They are potential candidates to contribute
to a new species
but they are considered as the inactive genes
for the species under consideration. If there happens another
individual with the mutant gene in the same locus
then these two mutants can mate (different sex is necessary as well
as the age $a \ge a_R$)
and they can produce offsrings.
In the latter case the new species is created with Eve who
 represents  Eve of the old species except for the mutant genes.
 In our
computer simulations the new species ususally
are extinct due to the mechanism of genetic drift. However, after long
periods of 'stasis' there happen bursts of the new species which are able
to live for a few thousands of generations or even more. They also can adapt
better to the surrounding and they can dominate other species.
The small changes of the inherited genetic information are realized
through the point mutations.

A point mutation changes a single bit within the 16-bit-string representation
of a gen and according to the above assumptions
the gene affected by a mutation may pass to one of three states:
$S=1$ (gene specific for the species),
$S=2$ (mutant gene, potential candidate of a new species),
$S=0$ (defected gene). In the model,
the bit-string representing a defected gene ($S=0$) can be mutated
with probability $p=0.1$. Hence, there is still a possibility for back mutations.
Genes are mutated only at the stage of the zygote
creation - one mutation per individual.

 The life cycle of diploids
needs an intermediate stage when one of the two copies of each gene
is passed from the parent to a haploid gamete. Next, the two gametes produced
by parents of different sex unite to form a zygote.

\begin{figure}
\includegraphics[height=6cm]{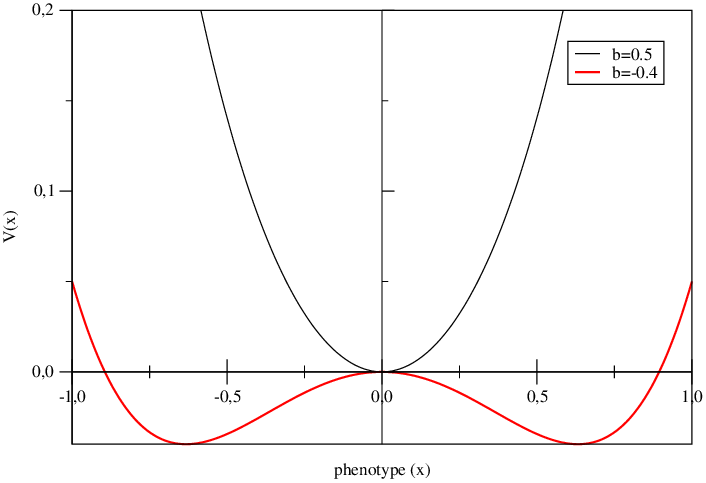}
\caption{Phenotype-surrounding interaction function
$V(x,t)=\frac{1}{2} x^4 + \frac{1}{2} b(t)   x^2$ for two values of $b(t)$
(0.5 and -0.4).
}
\label{fig1}
\end{figure}

In the model, phenotype is defined as a fractional representation of the
 16-bit-string genes, where each gene is translated uniquely into
a fractional number $x$ from the interval $[-1,1]$. The phenotype is coupled with
the surrounding with the help of the function $V(x,t)$ as follows:

\begin{equation}
V(x,t)=\frac{1}{4} x^4 + \frac{1}{2} b(t) x^2
\label{potencjal}
\end{equation}

\noindent
where the fluctuations of the surrounding are represented by the coefficient $b(t)$.
 The function
$V(x,t)$ has one minimum, $x=0$, for $b>0$ and two minima, $x=\pm \sqrt{|b|}$,
for $b<0$ as in Fig.\ref{fig1}. We use them to calculate
gene quality, $q$,
 in
the variable surrounding

\begin{equation}
q=e^{-(V(x,t)-V_{min}(t))/T}, ~~~~q \in [0,1],
\end{equation}

\noindent
where the parameter $T$ has been introduced
to control selection. The smaller value of $T$
the stronger selection.
Next, we calculate the fitness $Q$ of the individuals, say at age $a$,
to the surrounding with
the help of the average

\begin{equation}
Q=\frac{1}{L'+a} \sum_{i=1}^{L'+a} q_i
\label{rQ}
\end{equation}

\noindent
with

\begin{equation}
q_i=\max~\{q_i^{(1)} \delta(S_i^{(1)},1),q_i^{(2)} \delta(S_i^{(2)},1)\},
\end{equation}

\noindent
where the indices $(1)$ and $(2)$ denote,
respectively, the first allele and the second allele at locus
$i=1,2, \ldots, L'+a$ and $S_i=0,1,2$ represent their states.
The symbols $\delta()$ denote Kronecker delta.
The inactive alleles
and mutant genes do not contribute to $Q$ and always $Q \in [0,1]$.

We decided to determine the amount of possible offsprings by
projecting the fitness of the parents, $Q_F$ (female) and $Q_M$ (male),
onto the number of produced zygotes

\begin{equation}
N_{zygote}  = \max \{1, 10 \times \min (Q_{F}, Q_{M})\},
\label{r:gamete}
\end{equation}

\noindent
where at least one zygote is produced and the maximum number of zygotes
is equal to 10.
In the computer simulations, the zygotes are mutated, after they
are created, and they can be eliminated
if at least one housekeeping gene becomes inactive.  The individuals may reproduce
themselves after they reach the age $a=a_R$.
The genotypes of the parents who are better adapted to the surrounding
generate more offsprings. The mutants, if they happen (they posses
genes with the Hamming distance $H>1$ from Eve) cannot mate with the non-mutants.

We assume that the sex of  a diploid individual is determined with the help of
a random number generator at the moment when it is born
and it is unchanged during its life.

\section{Computer algorithm}

We investigate the species with respect to speciation. In most computer
simulations we observed evolving single species and analyzed the offsprings
of the new mutant species originating from the old one. The secondary
order speciation, i.e., the speciation taken place
from the mutant species, has not been considered.
However, we investigated the case when initially there
are a few species in the ecosystem and the results qualitatively were the same
as for single initial species.

In the simulation, it
is very important to prepare the genetic information in a proper way.
It is obvious that in the evolution process
some genes are not necessary during the
whole life, e.g., they may become important only near the end of life,
and it is possible that they could be defective since the individual
under consideration was born. Therefore, first we prepare the initial species
in the time independent surrounding for a few thousands generations until
the inherited genetic information is represented by  a steady state flow
between suceeding generations.
Then the distribution
of inactive genes becomes time independent. Only after that we switch
on the fluctuations of the surrounding, which in our case
are represented by the
coefficient $b(t)=-\frac{1}{2} - cos(\omega t)$, and all the initially prepared
species are put together in the ecosystem. Hence, the computer algorithm consists
of two blocks: preparation and evolution in variable surrounding.

\begin{itemize}
\item[]{\bf INITIAL SPECIES PREPARATION:}
   \begin{itemize}
     \item[(1)] generate $L$ bit-strings (16 bits) representing life functions
     of the organism
     and which are the patterns for genes
     \item[(2)] generate $N_s$ predecessors (Eve) of the initial species (Hamming
                distance of each gene from the respective pattern, $H<=2$,
                and then $S_1=1, S_2=1, \ldots S_L=1$)
     \item[(3)] for each Eve, representing the species $1,2, \ldots, N_s$
                construct $N_M$ males and $N_F$ females ($N_M=N_F$) who have
                 each gene
                at distance $H \le 1$ from the corresponding gene of Eve.
     \item[(4)] evolve separately each species $1,2, \ldots, N_s$ as follows:
     \begin{itemize}
      \item[(i)] increase the age of all individuals by 1
      \item[(ii)] remove from the population all individuals who should die because they have
               exceeded the maximum age ($L-L'$) or during their life until now
               they have collected
               three inactive "chronological" genes.
      \item[(iii)] [optionally] the Verhulst factor is applied, i.e.
               every individual survives with the same probability
               $(1-(N_M+N_F)/N_{max})$,
               where $N_{max}$ is the maximum allowed number of individuals
               in the population under consideration.
      \item[(iV)] select at random a female and a male
      for whom $a \ge a_R$ and let them
               produce $N_{zygotes}$ according to Eq.\ref{r:gamete}.
      \item[(V)] each zygote is mutated. We assume that
               the housekeeping genes cannot be inactive. Therefore, at least
               one defected housekeeping
               gene kills zygote. The total numer of the zygotes in the population
               which survive cannot exceed the number
               $\varepsilon_{pop} (N_E/\Omega) N_R$, where $N_R$ is the total number
               of pairs of individuals, female and male, whose age $a \ge a_R$.
               The parameter $\varepsilon_{pop}$ controls
               the allowed number of zygotes.
      \item[(Vi)] Decrease the total number of energy units by the number $N(t)$
      of all
      individuals (parents and their new-born offsprings), i.e., $N_E(t)=N_E(t-1)-N(t)$
      \item[(Vii)] regenerate the ecosystem energy according to Eq.\ref{rozw2}
      \item[(Viii)] goto (i) unless the stop criterion  is fulfilled
     \end{itemize}
   \end{itemize}
\item[] {\bf SPECIES IN VARIABLE SURROUNDING:}
\begin{itemize}
  \item[(5)] all $N_s$ species are put together in the ecosystem.
  \item[(6)] $t=t+1$
  \item[(7)] calculate $b(t)$ representing surrounding fluctuation
  \item[(8)] check all individuals from the old species with respect to mutants
  (who have, at least at one locus, $k=1, \ldots, L$, two bit-strings
           representing the pair of the
           alleles  at state $S^{(1)}=2$ and
           $S^{(2)}=2$).
           If there is in the population an individual representing a mutant
           then create Eve, the pattern of new species, and
           look for other mutants in the same species consistent with
           the created Eve. Determine a new
           value of the reproduction age $a_R \in [1,L-L']$
           for the species with the help of the computer random number generator.
         Remove new species from the old species.

  \item[(9)] step (i) from above
  \item[(10)] step (ii) from above
  \item[(11)] step (iii) from above
  \item[(12)] step (iV) from above
  \item[(13)] step (V) but now $N_R$ concerns all species.
  \item[(14)] step (Vi), where N(t) concerns all species
  \item[(15)] step (Vii)
  \item[(16)] goto (6) unless the end of the simulation
\end{itemize}
\end{itemize}

\section{Results and discussion}

Limiting the amount of energy resources in the ecosystem introduces
strong selection between the species competing for the resources.
In our case the selection takes place through limiting the number of
the new-born offsprings in the ecosystem. Namely, in the model,
their number cannot exceed
the value  $\varepsilon_{pop} (N_E(t)/\Omega) N_R$, where $N_R$
is the number
of parents, female and male,
from all species. Thus, the individuals
who are better adapted to the environment
(larger value of $Q$, Eq.\ref{rQ}) dominate other individuals because they
produce more offsprings.
In our model, the better adaptation results from the phenotype
$x$ closer to the minimum $x_{min}$ of the function $V(x,t)$.
We could expect that in a time independent
surrounding the mutations decreasing the fitness of the organism
lead to its elimination and in result to the
elimination of the affected genes.
However, in a changing
environment the species need an ability to react to new situations.
The quality of genes in a fluctuating surrounding is changing, i.e.,
we cannot tell any more that some genes are bad and some are good.
There will be preferred genes which adapt better to the changes.

\begin{figure}[ht]
\includegraphics[height=6cm]{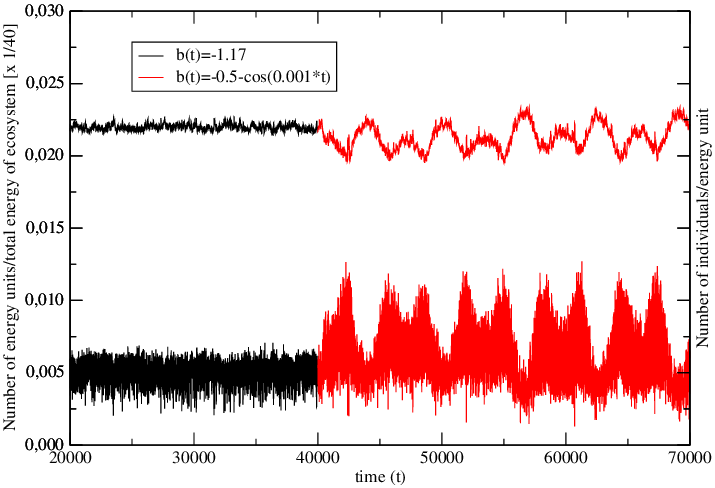}
\caption{ The effect of the transition from the time independent surrounding
($b(t)=-1.17$,  $t<40000$) to the one varying in time
($b(t)=-\frac{1}{2}-\cos(0.001 t $),  $t \ge 40000$)
on the size of the single
species ($a_R=5$, $a_{max}=11$)
and on the number of the energy units available for the species.
The bottoom curve follows the size changes of the species whereas
the upper curve shows the corresponding energy changes in the ecosystem.
}
\label{fig2}
\end{figure}

In Fig.\ref{fig2}, we show the effect of the changing environment on the
old population ($a_R=5$, $L=16$, $L'=5$), which was aging in a time independent
surrounding through 40000 generations. The changing enviroment is modelled with the
help of the coefficient $b(t)$ in the function $V(x,t)$ (Eq.\ref{potencjal}).
We have chosen

\begin{equation}
b(t)=-\frac{1}{2}-cos(\omega t),
\end{equation}

\noindent
where $\omega$ denotes the frequency of the environment changes.
We should take into account that although we have a single species,
in the example presented in Fig.\ref{fig2}, there are also present
energy resources in the ecosystem which regenerate themselves
according to Eq.\ref{rozw2}. Thus, we have a kind of the
Lotka-Volterra system with the species as the predator and the energy as the prey.
Once the
fluctuating environment changes fitness $Q$ of all individuals, the number of
the new-born offsprings follows the environment fluctuations and in
consequence there appear the energy oscillations. Notice, that in a changing
environment the size of the population under consideration undergoes much stronger
fluctuations than in the time-independent case. It is connected with the changes
of the inherited genetic information.

 \begin{figure}
\includegraphics[height=6cm]{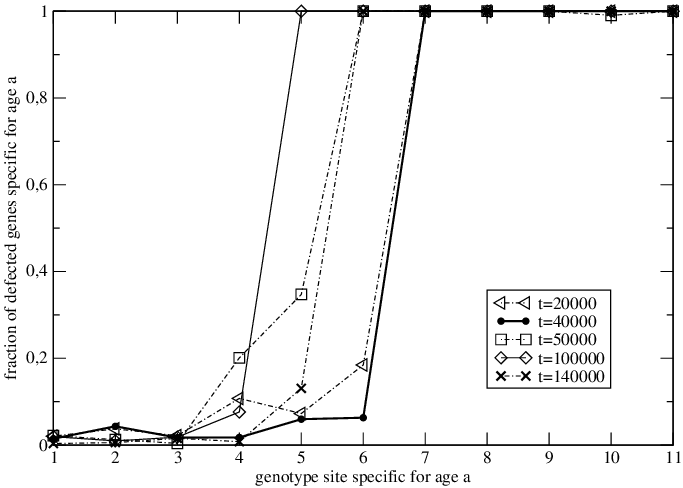}
\caption{ The effect of the change of the time independent surrounding
to the one varying in time, at $t \ge 40000$,
on the distribution of the defected "chronological"
genes for the situation presented in Fig.\ref{fig2}.
}
\label{fig3}
\end{figure}

\begin{figure}
\includegraphics[height=6cm]{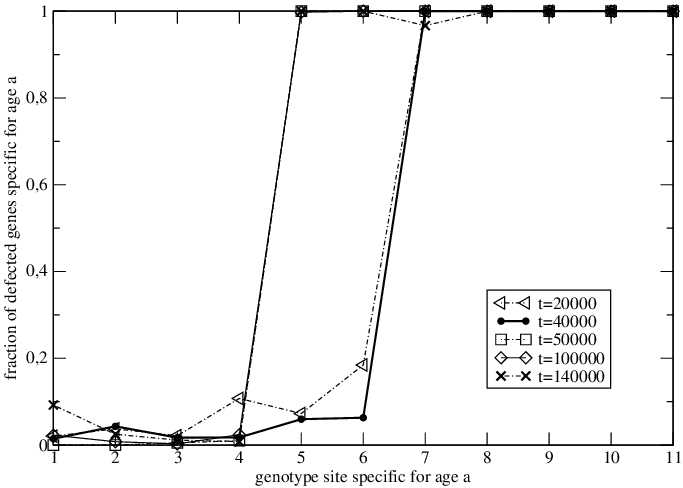}
\caption{ The same as in
 in Fig.\ref{fig2} but  $\omega=0.01$.
}
\label{fig4}
\end{figure}

In Fig.\ref{fig3} we have presented the age profiles of the  fraction of
the inactive "chronological" genes in the population from Fig.\ref{fig2}. The
profiles
concerning the generations at $t=20000$ and $t=40000$
refer to time independent surrounding.
Although we have introduced in the model the back mutations for
the inactive genes (the bit-strings corresponding to inactive genes may
mutate into another bit-string with probability $\rho=0.1$) we can observe
that, while time is increasing,
the shape of the profiles tends to the one which is time independent.
It determines the average life span
of individuals which is ranging from the age $a=1$ to $a=7$.
On the other hand, the profiles corresponding to time dependent
environment show both shrinking and enlarging of the life span.
Effectively, the
population becomes much younger than during its previous evolution
in the time independent surrounding. More frequent
fluctuations of the surrounding make this effect even stronger. It is evident
from Fig.\ref{fig4} where the same old species, under consideration,
experiences the
surrounding changing with the higher frequency $\omega=0.01$.
We can observe that the effect of the frequently
changing environment becomes more destructive
for the older members of the species. All our simulation were done with
a constant mutation rate (one mutation per genome) and we did not discuss
the effect of the various values of the mutation rate on the above results.

\begin{figure}
\includegraphics[height=6cm]{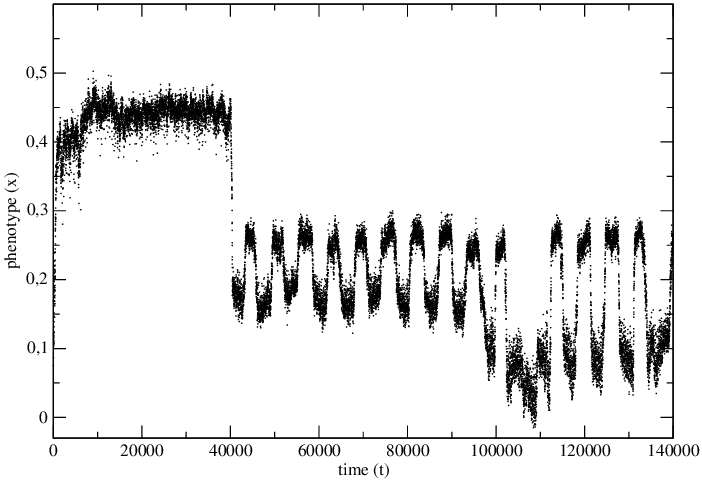}
\caption{
The phenotype evolution for the situation presented in Fig.\ref{fig2}.
}
\label{fig5}
\end{figure}

\begin{figure}[ht]
\includegraphics[height=6cm]{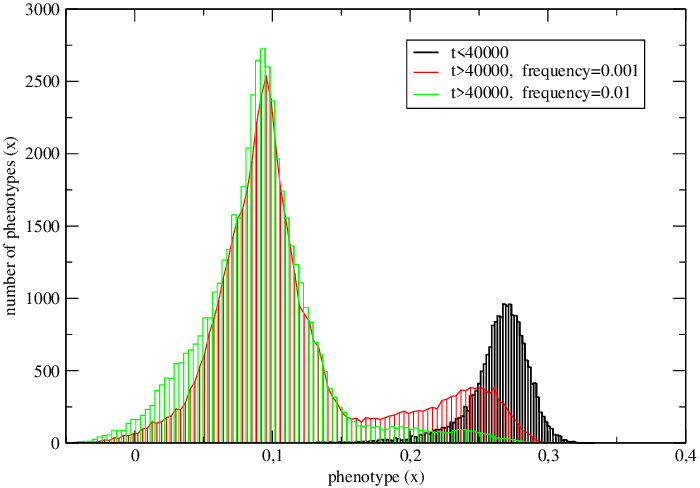}
\caption{The histogram of the average usage of the phenotype $x$
in the evolving species (from Fig.\ref{fig2})
when there is time independent surrounding (r.h.m histogram),
and when there is fluctuationg environment.
}
\label{fig6}
\end{figure}

The evolution of the average phenotype of the species from the example
in Fig.\ref{fig2} has been shown in Fig.\ref{fig5}. The same data, but
presented with the help of a histogram have been shown in Fig.\ref{fig6}.

\begin{figure}
\includegraphics[height=6cm]{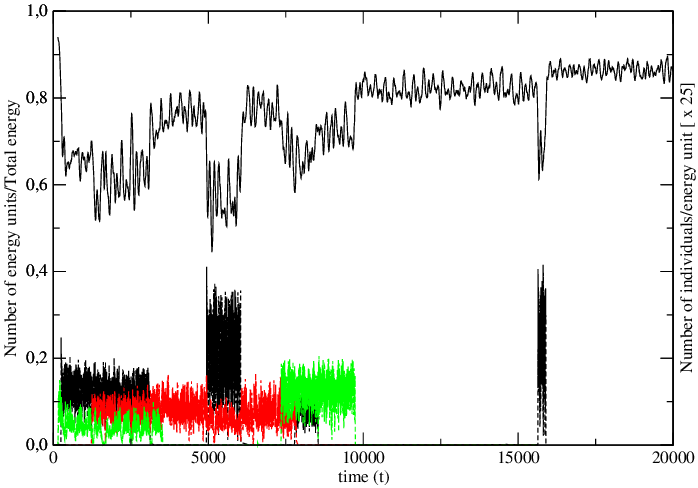}
\caption{ The effect of the speciation from the old species
(earlier
aging for 40000 generations in time independent environment)
in a fluctuating environment ($b(t)=-\frac{1}{2}-\cos(0.01 t $) on the
evolution of the energy resources.
In the bottom part of the figure there have been shown a few examples
of the new species created in the course of the evolution. The energy
cost of the new species creation or extinction is reflected by
the terrace-like changes in the upper curve. The survived species
have not been shown for clarity of the picture.
}
\label{fig7}
\end{figure}

\begin{figure}
\includegraphics[height=6cm]{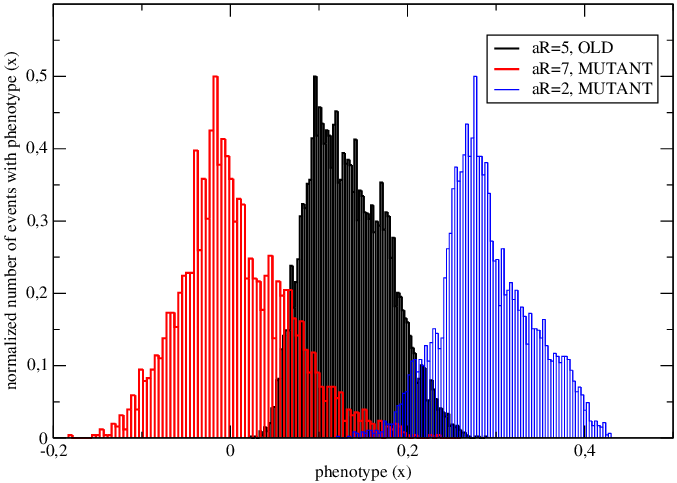}
\caption{Histograms of the phenotype usage in the environment
changing with the frequency $\omega=0.02$ in the case of
three species: the predecessor population ($a_R=5$),
and two descendant populations ($a_R=7$,$a_R=2$). The frequency
of the environment changes, $\omega=0.02$.
}
\label{fig8}
\end{figure}

Notice, that there is a trend to minimize the  environment fluctuations
and the genes, which were well adapted in the time independent environment,
 do not have the ability to follow the environment fluctuations. They can be even
 eliminated from the genetic bank of the species if the fluctuations become
 more frequent.

 The environmental changes influence the size of all species in the ecosystem
 end this is the reason that
 some of them may be eliminated. The regeneration of the energy resources
 in the ecosystem takes some time and too rapid a growth of some species, e.g.,
 mutant species,
 acts as a sudden cataclysm for the other species.
 The Lotka-Volterra systems have a self-regulatory character and there exist
 threshold values for the fraction of destroyed population above which the system
 returns to its previous state (\cite{katastrofa},\cite{Artur}).
 However, in the case of genetic populations, if the catastrophe is
 applied for many generations, one can observe that
 the age profile of the fraction of defective
 genes in the population may loose its stability and next
 the species becomes extincted  (\cite{Artur}).
Thus the appearance of the new species, even
 for a relatively short period, can eliminate the old species.

 In our computer simulations we
 associate a new value, $a_R$, of the reproduction age with every new species
 appearing in the course of evolution. The value is determined with
 the help of a computer random number generator. Thus, we have also
 the possibility to investigate the effect of the reproduction age on the
 adaptation of the species to the fluctuating surrounding. We have
 observed that the
 events, representing  the appearance of the new species
with the small value $a_R$ ($a_R \sim 1$), are usually
represented by short duration
 "bursts" of the population size and
 they vanish as rapidly as they have appeared. In the example from
 Fig.\ref{fig7} the are
 represented by the highest blobs painted in black, at the bottoom
  of the figure.
 It is easy to explain the phenomenon, as in this case the life span
 of individuals
 practically  is shrinking to the activity of single gene and the individuals
 die after they reproduce themselves. The active genes cannot effectively
 adapt to the changing environment. How important the range of life span for
 species stability is, can be also concluded from the histograms in
Fig.\ref{fig8}.

There are presented histograms of the average usage of the phenotype $x$
of the species representing the predecessor ($a_R=5$) and two descendant
species ($a_R=2,7$). It is evident that the average value $x$
representing the species with $a_R=2$ oscillates far away from the optimum
values (variable minima of $V(x,t)$), because there is an insufficient number
of genes adapted to the variable surrounding. The situation
is a little bit better with the
predecessor (old species). However, it is evident that it cannot be stable
over a long time in a variable surrounding.
The most stable species is the one for which $a_R=7$,
because in its population there are inherited genes which have adapted
to the variable surrounding
(the range of $x$ is almost symmetric with respect to $x=0$).

In this study,
 we did not discuss
 the mechanism of speciation in the new species. We have restricted
 our analysis to
  the "first order" speciation originating from old species and we discussed
 in detail its effect on the inherited genetic information.
We could expect a chain of events representing speciation
with the species being better and better adapted to the environment.
The younger the species the more probable speciation, and one should observe
peaks in the number of the speciation events in a short period.

\section{Conclusions}
We have discussed a model of species evolution in an ecosystem
where the energy resources regenerate themselves according to a logistic
differential equation. The model belongs to  the class of the Lotka-Volterra systems
where the energy resources represent prey and the species represent predator.
The number of the species present in the ecosystem under consideration
results from evolution,
which is understoood as the interplay of the two processes only, mutation
and selection. We observed that after long periods of
'stasis' there happen bursts of the new species which are able
to live for thousands of generations.
We have observed that in a variable surrounding
the species, for which the reproduction age $a_R$ is too small, they
are very unstable even if their size substantially exceeds the size of
the populations specific for other species. In our model, they usually cause
the elimination of other species. Simultaneously, the resulting increase in
the energy resources  makes  the next speciations possible.

There are two
approaches to the description of species evolution: the one with
continuous evolutionary changes and the theory of
punctated equilibrium \cite{SOC1,SOC2},
according to which the evolutionary activity occurs in bursts.
It is often the case that the mathematical models concerning the species evolution
are restricted to pure species dynamics consideration or pure genetics evolution.
We have shown
that the inclusion of the genetic information into the population dynamics
makes the use of  Bak-Sneppen \cite{SOC1,SOC2} extremal dynamics (SOC) possible.

\end{document}